# Pseudofermion observables for static heavy meson decay constants on the lattice

G. M. de Divitiis, R. Frezzotti, M. Masetti and R. Petronzio

Dipartimento di Fisica, Università di Roma *Tor Vergata*

and

INFN, Sezione di Roma II

Viale della Ricerca Scientifica, 00133 Roma, Italy

March 22, 1996

## Abstract

A method based on the Monte Carlo inversion of the Dirac operator on the lattice provides low noise results for the correlations entering the definition of the heavy meson decay constant in the static limit. The method is complementary to the usual method of smeared sources, avoids the systematic error arising from optimizing the size of the smearing volume and is more efficient for the values of lattice parameters that we have explored.

The knowledge of leptonic decay constants of heavy mesons is crucial for the extraction of the Cabibbo-Kobayashi-Maskawa mixing angles. In particular the value of the $B$ meson decay constant constraints the shape of the unitarity triangle and sets the size of possible CP violation in $B$ decays. Lattice calculations are attempting since a few years to provide non perturbative estimates of this quantity, but are still confronted with various sources of systematic errors. One of them originates from the presence in the dynamics of a heavy meson system of two very different mass scales. Reasonable values for exploring the $B$ meson dynamics with enough resolution of the heavy quark propagation and enough volume for the heavy meson wave function are a lattice size of the order of 1.5 fm and a lattice spacing of order of 0.02 fm which lead to about 75 lattice points, a value beyond present computer capabilities.

One reaches the physical region of the heavy quark mass by interpolating between the results which are obtained in the charm quark mass region and those deriving from an expansion of the fermion action in the inverse of the heavy quark mass. The first term in the latter expansion is the so called static limit [1] and corresponds to the approximation where the heavy quark does not propagate in space. An accurate knowledge of the decay constant in this limit is an essential ingredient of the calculation of the physical value. In spite of the great computer effort dedicated to its estimate the values quoted in recent papers are still affected by a $10-15\%$ error, an important fraction of which is due to the large statistical fluctuations. In this letter we propose an alternative way of measuring the correlator of the heavy meson in the static limit which makes possible to reduce the statistical errors significantly.

On the lattice and in the static limit, the correlator of two local heavy-light pseudoscalar bilinears is given by:

$$G(t) = \frac{1}{L^3 T} \sum_{x,y} \delta^3_{\vec{x},\vec{y}} \delta_{t_y, t_x+t} \langle \text{Tr}[P(x,y) \frac{1-\gamma_0}{2} S_q^\dagger(x,y)] \rangle \qquad (1)$$

where $S_q$ is the light quark propagator, $P(x,y)$ is the Polyakov line from the point $x$ to $y$, and $L$ and $T$ are the space and time sizes respectively.

At large times there is a single state dominating the spectral decomposition of the correlation function and one can extract the matrix element of the heavy-light current between such a state and the vacuum:

$$G(t) \stackrel{t \; large}{\longrightarrow} Z_L^2 e^{-\Delta E t} \tag{2}$$

The decay constant in the static limit can be derived from $Z_L$:

$$f_P \sqrt{M_P} = \sqrt{2} Z^{Ren} Z_L a^{-3/2} \tag{3}$$

where $Z^{Ren}$ is the suitable renormalization constant of the lattice current.

With standard methods the signal of the local currents is too noisy to wait for large times where the lowest lying state dominates : the extraction of the matrix element is then performed in two steps. In the first one uses smeared currents which optimize the projection on the lowest mass state and allow an early estimate of the energy shift in eq. 2, in the second, one analyses the ratio of single and double smeared correlation functions to get the estimate of the local matrix element. Error propagation is minimized by performing fits of these correlations globally. A typical source of systematic error arises from the size of the smearing volume and from the choice of the window in time used for the fits.

In the static approximation, the final space point of the correlator coincides with the initial space point. The inversion of the Dirac operator is very expensive in computer time and is normally performed for a single initial point on each gauge configuration. The major source of statistical errors can be ascribed to the difficulty of estimating the correlator in eq. 1 for many independent initial points.

The method presented in this letter is based on a Monte Carlo inversion of the Dirac operator which allows to perform the average of the static correlator over all the initial points, to obtain a visible signal at large enough times and to avoid the use of smeared sources. We introduce an auxiliary action with pseudofermion fields:

$$S_\phi[U, \phi] = \sum_x |[Q\phi](x)|^2 \tag{4}$$

where $\phi(x)$ is the pseudofermion field, and for the lattice Dirac operator $D$ we follow the standard Wilson formulation:

$$[Q\phi](x) = \gamma_5 D \phi = \frac{1}{2\kappa} \gamma_5 \phi(x) - \frac{1}{2} \gamma_5 \sum_{\mu=0}^{3} U_\mu(x)(1 - \gamma_\mu)\phi(x + \mu)$$

$$- \frac{1}{2} \gamma_5 \sum_{\mu=0}^{3} U_\mu^\dagger(x - \mu)(1 + \gamma_\mu)\phi(x - \mu) \tag{5}$$

where $\kappa$ is the Wilson hopping parameter related to the bare mass.

The sum $S_g[U] + S_\phi[U, \phi]$, where $S_g$ is the standard Wilson action for the gauge sector, is the "bermion" action used in the study of the dynamical flavour dependence of lattice QCD by extrapolating from negative flavour numbers [2]. In the context of this letter, the bermion fields are not dynamical, they are thermalized in a fixed gauge confuguration and only used to estimate fermion propagators, like in the fermion algorithm of ref. [3]. The presence of the square of the Dirac operator is needed for the Montecarlo update of the pseudofermion fields and implies that the two point function of these fields represents the inverse of such a square. The usual inverse Dirac operator can be obtained by the remultiplying the pseudofermion field correlations with the Dirac operator. We define two correlators, the first corresponding to eq. 1:

$$G(t) = \frac{1}{L^3 T} \sum_{x,y} \delta^3_{\vec{x},\vec{y}} \delta_{t_y, t_x + t} \langle [Q\phi]^\dagger(x) \gamma_5 P(x,y) \frac{1+\gamma_0}{2} \phi(y) \rangle \qquad (6)$$

and a second one corresponding to propagators with the square of the Dirac operator:

$$G^{(Q^2)}(t) = \frac{1}{L^3 T} \sum_{x,y} \delta^3_{\vec{x},\vec{y}} \delta_{t_y, t_x + t} \langle \phi^\dagger(x) P(x,y) \frac{1+\gamma_0}{2} \phi(y) \rangle \qquad (7)$$

The statistical fluctuations of these operators can be reduced by a "one $\phi$ integral" technique analogous to the "one link integral" method [5]: one replaces the pseudofermion field by its average in the surrounding pseudofermion and gauge configuration. The explicit expression for such a replacement is:

$$\phi(x) \to -A^{-1}[Q^2 \phi - A\phi](x) \qquad (8)$$

where

$$A = \frac{1 + 16\kappa^2}{4\kappa^2} \qquad (9)$$

The action for the fields which are replaced in the same observable through eq. 8 should be separable, which implies that the improved correlation functions $G(t)$ and $G^{(Q^2)}(t)$ can only be calculated at time distances greater than two.

As already noticed earlier [4] the correlator constructed from the inverse of the square of the Dirac operator in eq. 7 projects more precociously on the lowest lying state. We have therefore extracted the energy shift $\Delta E$ from this correlator and then fixed its value in the fit of the canonical correlator $G(t)$. We have tested this method at two values of $\beta$: 5.7 and 6.0 on a $16^3 \times 32$ lattice by measuring correlation functions in all four directions. The simulations were performed on a 25 Gigaflop machine of the APE series. The update procedure was for the gauge sector a Cabibbo–Marinari pseudo–heatbath [6] followed by three overrelaxation sweeps and for the pseudofermions a heat bath followed by ten overrelaxation sweeps [7]. The gauge field configurations were separated by 1000 sweeps and for each fixed gauge configuration the calculation of the pseudofermion observables was performed every 5 pseudofermion sweeps in a total of 2000 pseudofermion updates, i. e. 400 times. In addition, for each gauge configuration, pseudofermion measurements were taken after 500 thermalization updates which were tested to be sufficient to equilibrate the pseudofermion system. The Montecarlo inversion of the Dirac operator with 400 samplings achieves a modest precision with respect to the inversion by a minimization algorithm, but becomes of comparable quality when one uses the sum over all space time points. For example, the zero momentum pion correlation, averaged over 30 gauge configurations, at $\kappa = 0.165$ has a 10% error at $t = 13$ if calculated with the Dirac deterministic inversion, and the same error at $t = 10$ if calculated with the Monte Carlo inversion. The sum over all space time points in the case of $f_B$ provides the additional advantage of sampling also the heavy quark propagator through the whole lattice. The computer time required with the 2000 updates and 500 thermalization sweeps of the pseudofermion system is comparable to that of a value of a single origin Dirac inversion at $m_\pi^2/m_\rho^2 \sim 0.4$.

The results of our simulation refer to a set of 30 gauge configurations. The errors are evaluated as follows: we form clusters of the 400 measurements of the correlation functions at fixed gauge configurations or larger clusters containing 3 gauge configurations, i. e. of a total of $400 \times 3$ measurements. This subdivides the sample of $30 \times 400$ pseudofermion measurements into 30 or 10 clusters respectively. On these clusters we apply a jacknife algorithm to estimate the errors of the correlations. Each "jacknife cluster", defined by single elimination of a standard cluster from the total sample is fitted with a MINUIT minimization program. From the spread of the fits on each jacknife cluster we extract the error that we quote for our results. The procedure with standard clusters gives consistent results within the quoted errors.

The results for the effective mass in the case $\beta = 5.7$ at a value of the hopping parameter $\kappa = 0.165$ corresponding to a ratio $m_\pi^2/m_\rho^2$ of 0.5 are presented in figure 1 and compared with the fit. For the operator of eq. 7 a one mass fit is generally adequate. For the operator of eq. 6 a two mass fit is necessary with the lowest mass value fixed by the fit of the other operator. The results for different values of $\kappa$ and for different choices of the starting value of the time window used in the fit are given in Table 1. In the case of the single mass fit the window is shifted to larger times to reach a better projection on the lowest mass state.

For $Z_L$ we give separately the statistical error of the $G(t)$ correlation fit and the error deriving from the uncertainty by which the correlation function $G^{(Q^2)}(t)$ fixes the lowest energy shift value. Our results can be compared with those of [8], obtained on a smaller volume ($12^3 \times 24$). Our values, less affected by finite size effects, lie in general below, exhibit smaller errors and are obtained with a third of the number of gauge configurations used in [8].

For the results at $\beta = 6.0$ one needs a two mass fit also for the correlation function $G^{(Q^2)}(t)$ as it can be seen from figure 2 which presents the effective mass at a value of the hopping parameter $\kappa$ corresponding to a ratio $m_\pi^2/m_\rho^2$ of 0.5. The continuous line is our two mass fit. Indeed, at $\beta = 6.0$ the lattice spacing is smaller and at fixed time distance in lattice units the correlation functions are sampled at smaller physical units where the excited states are still important. The compensating effect of a smaller energy shift in lattice units which would allow the signal to survive up to larger lattice times is partly lost because of the linear divergences which contribute with a constant to the mass in lattice units. This effect is stronger in the correlation of the operator of eq. 6: in this case, the results for $Z_L$ of the two mass fit show a systematic decrease with increasing values of the starting time of the window used for the fit. This does not happen for the results at $\beta = 5.7$ which are insensitive to the choice of the fitting window. The final value we quote is obtained from the last time window, but should still be considered as an upper bound of the true value. The new preliminary results of the APE [9] and MILC [10] collaborations at $\beta = 6.0$ are indeed compatible with ours but lie below within one standard deviation. The extrapolation of our results to the chiral limit interpolates rather well between the results at $\beta = 6.1$ and $\beta = 5.9$ of [8]. A work applying this method to the study of the dynamical flavours dependence of $f_B$ where the pseudofermions are replaced by the dynamical bermion fields is in progress [11].

Our method appears to reach a better statistical precision with respect to standard method. The gain in statistics comes from three sources: the Montecarlo Dirac inversion which provides an alternative way of measuring correlations where, as in the case of static pseudoscalar decay constant, the additional average over all points is important, the measurement of a correlator which leads to an earlier determination of the lowest lying state and the "one $\phi$" integral which reduces the statistical fluctuations. The method discussed in this paper is complementary to the existing smearing procedures and can be conjugated with them to further reduce the statistical and systematic uncertainties of lattice determination of static pseudoscalar decay constants.

**Acknowledgments.** We thank C. Allton and C. Bernard for communicating to us their preliminary results prior to publication and A. Vladikas for stimulating discussions.

# References

[1] E. Eichten, B. Hill, Phys. Lett. B 232 (1989) 113.

[2] R. Petronzio, Nucl. Phys. B (Proc. Suppl.) 42 (1995) 942; G. M. de Divitiis, R. Frezzotti, M. Guagnelli, M. Masetti, R. Petronzio, Nucl. Phys. B 455 (1995) 274; G. M. de Divitiis, R. Frezzotti, M. Guagnelli, M. Masetti, R. Petronzio, Ph. Lett. B 367 (1996) 279.

[3] F. Fucito, E. Marinari, G. Parisi, C. Rebbi, Nucl. Phys. B 180 (1981) 369.

[4] G. M. de Divitiis, R. Frezzotti, M. Guagnelli, M. Masetti, R. Petronzio, Ph. Lett. B 353 (1995) 274.

[5] G. Parisi, R. Petronzio, C. Rapuano, Phys. Lett. B 128 (1983) 418.

[6] N. Cabibbo, E. Marinari, Ph. Lett. B 119 (1982) 387.

[7] M. Lüscher, Nucl. Phys. B 418 (1994) 637; B. Bunk, K. Jansen, B. Jegerlenner, M. Lüscher, H. Simma, R. Sommer, Nucl. Phys. B (Proc. Suppl.) 42 (1995) 49.

[8] A. Duncan, E. Eichten, J. Flynn, B. Hill, G. Hockney, H. Thacker, Phys. Rev. D 51 (1995) 5101.

[9] C. Allton, private comunication.

[10] C. Bernard, private comunication.

[11] G. M. de Divitiis, R. Frezzotti, M. Masetti, R. Petronzio, in preparation.

| $\kappa$ | $\Delta E(t_{\min} = 4)$ | $\Delta E(t_{\min} = 5)$ | $\Delta E(t_{\min} = 6)$ | $\Delta E$ |
|---|---|---|---|---|
| 0.163 | 0.8145(6) | 0.8126(7) | 0.8118(10) | 0.812(2) |
| 0.165 | 0.7848(7) | 0.7853(9) | 0.7865(12) | 0.786(2) |
| 0.16625 | 0.7645(9) | 0.7671(12) | 0.7701(20) | 0.770(4) |
| $\kappa$ | $Z_L(t_{\min} = 3)$ | $Z_L(t_{\min} = 4)$ | $Z_L(t_{\min} = 5)$ | $Z_L$ |
| 0.163 | 0.649(4) | 0.655(8) | 0.655(20) | 0.655(8)[7] |
| 0.165 | 0.590(7) | 0.591(10) | 0.590(20) | 0.590(7)[7] |
| 0.16625 | 0.558(7) | 0.559(9) | 0.560(20) | 0.559(9)[12] |

Table 1: Results for $\Delta E$ and $Z_L$ calculated on a $16^3 \times 32$ lattice at $\beta = 5.7$. In the last column we quote our final result: for $Z_L$ we separate the statistical error () from the one coming from the determination of $\Delta E$ [].

| $\kappa$ | $\Delta E(t_{\min} = 3)$ | $\Delta E(t_{\min} = 4)$ | $\Delta E(t_{\min} = 5)$ | $\Delta E$ |
|---|---|---|---|---|
| 0.154 | 0.635(7) | 0.635(10) | 0.632(30) | 0.635(7) |
| 0.155 | 0.616(7) | 0.616(10) | 0.615(18) | 0.616(7) |
| 0.1557 | 0.598(10) | 0.598(14) | 0.598(20) | 0.598(10) |
| $\kappa$ | $Z_L(t_{\min} = 3)$ | $Z_L(t_{\min} = 4)$ | $Z_L(t_{\min} = 5)$ | $Z_L$ |
| 0.154 | 0.250(4) | 0.244(5) | 0.237(8) | 0.237(8)[10] |
| 0.155 | 0.231(3) | 0.223(5) | 0.220(7) | 0.220(7)[7] |
| 0.1557 | 0.213(3) | 0.207(5) | 0.200(7) | 0.200(7)[10] |

Table 2: The same as in table 1 for $\beta = 6.0$.

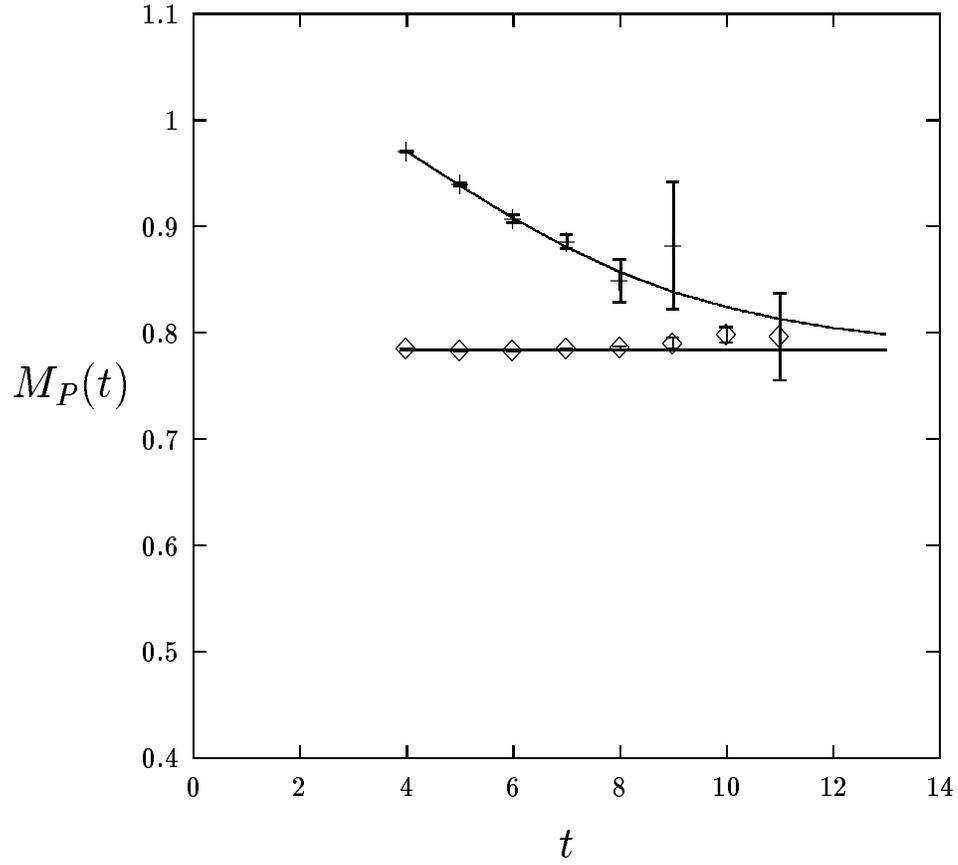

Figure 1: The effective mass for the correlation function $G^{(Q^2)}(t)$ (diamonds) is compared with the single mass fit and the effective mass for the correlation function $G(t)$ (crosses) is compared with the two mass fit for $\beta = 5.7$ and $\kappa = 0.165$.

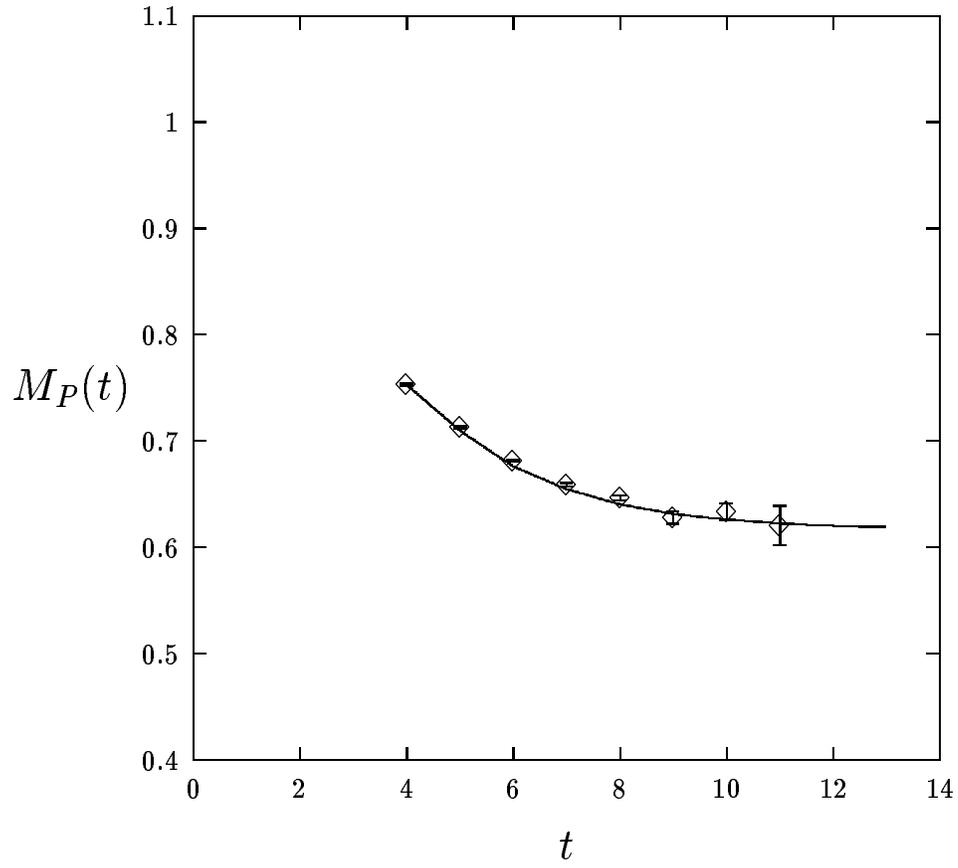

Figure 2: The effective mass for the correlation $G^{(Q^2)}(t)$ for $\beta = 6.0$ and $\kappa = 0.155$ is compared with the two mass fit.